\documentclass[a4paper,fleqn]{cas-dc}

\usepackage[authoryear]{natbib}

\usepackage{amsmath,amsfonts,amssymb}

\usepackage{graphicx}

\usepackage{amsthm}

\usepackage{multirow}
\usepackage{booktabs}
\usepackage{multicol}

\usepackage{subcaption}
\usepackage{caption}

\usepackage{array}
\usepackage{adjustbox}
\usepackage{changepage}
\usepackage{tabularx}
\def\tsc#1{\csdef{#1}{\textsc{\lowercase{#1}}\xspace}}
\tsc{WGM}
\tsc{QE}
\tsc{EP}
\tsc{PMS}
\tsc{BEC}
\tsc{DE}

\newcommand{\con}{{\,\vert \,}}

\newcommand{\half}{{\frac{1}{2}}}

\newcommand{\tausq}{{\tau^2}}

\theoremstyle{plain}

\theoremstyle{remark}

\begin{document}
\let\WriteBookmarks\relax
\def\floatpagepagefraction{1}
\def\textpagefraction{.001}
\shorttitle{}
\shortauthors{Datta, Shudde, Johnson}

\title [mode = title]{Bayes Factors Based on Test Statistics and Non-Local Moment Prior Densities}                      

\tnotetext[1]{This document is the result of the research project funded by the National Science Foundation [DMS 2311005].}

\author[1]{Saptati Datta}[
                        orcid=0009-0009-3331-6523]

\credit{Conceptualization, Formal Analysis, Methodology, Software, Writing - original draft and revision}

\affiliation[1]{organization={Texas A\&M University},
                addressline={TAMU}, 
                postcode={77843}, 
                postcodesep={}, 
                city={College Station},
                country={USA}}

\author[1]{Rachael Shudde}[orcid=0000-0003-1672-3118]
\credit{Methodology, Software}

\author[1]{Valen E. Johnson}[orcid=0000-0002-8659-4772]
\cormark[1]
\ead{vejohnson@tamu.edu}

\credit{Conceptualization, Formal Analysis, Methodology, Software, Writing - original draft and revision, Supervision}



\begin{abstract}
   We describe Bayes factors based on \emph{z}, \emph{t}, $\chi^2$, and \emph{F} statistics when non-local moment prior distributions are used to define alternative hypotheses. The non-local alternative prior distributions are centered on standardized effects. The prior densities include a dispersion parameter that can be used to model prior precision and the variation of effect sizes across replicated experiments. We examine the convergence rates of Bayes factors under true null and true alternative hypotheses and show how these Bayes factors can be used to construct Bayes factor functions. An example illustrates the application of resulting Bayes factors to psychological experiments.
\end{abstract}


\maketitle

\section{Introduction}

Bayes factors based on test statistics and first-order non-local prior densities were used in \cite{Johnson2023} (hereafter J23) to define Bayes factor functions (BFFs).  The first-order non-local prior densities used to define those Bayes factors contained a single scale parameter that determined the mode of the non-local prior densities used to define the alternative hypotheses \citep{Johnson2010}. BFFs were defined as the mapping of these prior modes to Bayes factors. The use of first-order moment prior densities allowed J23 to obtain closed-form expressions for Bayes factors based on common test statistics, including $z$, $t$, $F$, and $\chi^2$ statistics. 

In this article we extend the results of J23 by providing closed-form expressions for Bayes factors based on test statistics (BFBOTs) \citep{Johnson2005} and alternative hypotheses defined using non-local moment prior densities of arbitrary order.
This extension enables the incorporation of subjective prior knowledge regarding the precision of prior estimates of non-null effect sizes. Moreover, it provides a potential mechanism for modeling variation in effect sizes across replicated experiments.

\section[Bayes factors for z, t, $\texorpdfstring{\chi^2}{chi^2}$, and F]{BFBOTs and moment prior alternatives} \label{sec:2}

J23 defined Bayes factors using two categories of prior densities: normal-moment prior densities and gamma prior densities.  A normal-moment density on a parameter $\lambda$, given hyperparameters $(\tau^2,r)$, can be expressed as 
\begin{eqnarray}\label{rnm}
j(\lambda\mid\tau^2, r) &=& \frac{(\lambda^2)^r}{(2\tau^2)^{r+\frac{1}{2}}\Gamma\left(r+\frac{1}{2}\right)}\exp\left(-\frac{\lambda^2}{2\tau^2}\right), \nonumber \\ 
& &  -\infty <\lambda<\infty,  \quad \tau,r>0.
\end{eqnarray}
J23 derived Bayes factors based on $t$ and $z$ statistics by imposing normal-moment prior densities with $r=1$ on the non-centrality parameters of the test statistics under the alternative hypothesis.  We extend these results for $r\ge 0$.
 
BFBOTs for $\chi^2_k$ and $F_{k,m}$ test statistics were defined by assuming the non-centrality parameter $\lambda$ for each distribution under the alternative hypothesis followed gamma distributions with parameters $k/2+1$ and $1/(2\tau^2)$, i.e., $G[k/2+1,1/(2\tau^2)]$. We generalize these results by extending the class of gamma priors to include $G[k/2+r,1/(2\tau^2)]$ distributions for $r,\tau>0$.

For one-sided tests with positive non-centrality parameters, the prior densities \( j^+(\mu \mid \tau^2, r) \) and \( j^-(\mu \mid \tau^2, r) \) are defined similarly to the expression in (\ref{rnm}), but they are constrained to the positive or negative real line, respectively. This constraint depends on whether the alternative hypothesis necessitates that \(\lambda > 0\) or \(\lambda < 0\).

For simplicity we adopt the notation used in J23 and define
\begin{itemize}
\item $N(a,b)$ as a normal distribution with mean $a$ and variance $b$.
\item $T_\nu(\lambda)$ as a $t$ distribution with $\nu$ degrees of freedom and  non-centrality parameter $\lambda$.
\item $\chi^2_\nu(\lambda)$ as a $\chi^2$ distribution with $\nu$ degrees of freedom and non-centrality parameter $\lambda$.
\item $F_{k,m}(\lambda)$ as an $F$ distribution with $(k,m)$ degrees of freedom and non-centrality parameter $\lambda$.
\item $G(\alpha,\lambda)$ as a gamma distribution with shape parameter $\alpha$ and rate parameter $\lambda$.
\item $J(\tau^2,r)$ as a normal-moment distribution of order $r$ and rate parameter $\tau^2$ and density given in~(\ref{rnm}).
\end{itemize}
In addition, we let ${_1}F_1(a,b;z)$ denote the confluent hypergeometric function and ${_2}F_1(a,b,c; z)$ the Gaussian hypergeometric function.

\begin{table*}[width=1.9\linewidth,cols=5,pos=t]
\caption{Summary of theorems describing Bayes factors based on test statistics. The constants $a$, $b$ and $c$ are defined as $a = \left(1+\tau^2\right)^{-\left(r+\half\right)}$, $b = \left(1+\tau^2\right)^{-\left(\frac{k}{2}+r\right)}$, and $c=\left[\Gamma(\frac{\nu}{2}+1)\,\Gamma(r+1)\right]/\left[\Gamma(\frac{\nu+1}{2})\,{\Gamma(r+\half)}\right] $.}\label{tbl1}
\begin{tabular*}{\tblwidth}{@{} LCCCC@{} }
\toprule
Test & $H_0$ & $H_1$ & Prior & $BF_{10}$ \\ \midrule
    Two-sided z  & $z\sim N(0,1)$ & $z\sim N(\lambda,1)$ & $J(\lambda\con\tausq,r)$ & $a\,F_1\left(r+\half,\half;
    \frac{\tau^2 z^2}{2(1+\tau^2)}\right)$ \\
     & & & & \\ 
    One-sided z & $z\sim N(0,1)$ & $z\sim N(\lambda,1)$ & $J^+(\lambda\con\tausq,r)$ & 
    $a\, \left[ {_1}F_1\left(r+\half,\half;y^2 \right) + 2 y \frac{\Gamma(r+1)}{\Gamma(r+\half)} {_1}F_1\left(r+\half,\frac{3}{2};y^2 \right) \right]$, \\
     & & & & \hfill with $ y = \tau z (2+2\tau^2)^{-\half}$ \\
    Two-sided $t$ & $t\sim T_\nu(0,1)$ & $t\sim T_\nu(\lambda,1)$ & $J(\lambda\con\tausq,r)$ &
    $a\, \left[ {_2}F_1\left(\frac{\nu+1}{2},r+\half,\half;y^2 \right) +
    cy\ \,   {_2}F_1\left(\frac{\nu}{2}+1,r+1,\frac{3}{2};y^2 \right)  \right]$ \\
    & & & & 
    \hfill with $ y = \tau t \,[(\nu+t^2)(1+\tau^2)]^{-\half}$ \\
    One-sided $t$ & $t\sim T_\nu(0,1)$ & $t\sim T_\nu(\lambda,1)$ & $J^+(\lambda\con\tausq,r)$ &
    $a\, \left[ {_2}F_1\left(\frac{\nu+1}{2},r+\half,\half;y^2 \right) + 
     2cy\ \,  {_2}F_1\left(\frac{\nu}{2}+1,r+1,\frac{3}{2};y^2 \right)\right] $ \\
    & & & & 
    \hfill with $ y = \tau t \, [(\nu+t^2)(1+\tau^2)]^{-\half}$ \\
    $\chi^2_k$ & $h \sim \chi^2_k(0)$ & $h\sim \chi_k^2(\lambda)$ & $G\left(\lambda\con\frac{k}{2}+r,\frac{1}{2\tau^2}\right)$ & 
    $b\ \, {_1}F_1\left(\frac{k}{2}+r,\frac{k}{2};
    \frac{\tau^2 h}{2(1+\tau^2)}\right)$ \\
    $F_{k,m}$ & $f \sim F_{k,m}(0)$ & $f\sim F_{k,m}(\lambda)$ & $G\left(\lambda\con\frac{k}{2}+r,\frac{1}{2\tau^2}\right)$ & 
    $b\ \, {_2}F_1\left(\frac{k}{2}+r,\frac{k+m}{2},\frac{k}{2};
    \frac{k\tau^2 f}{(1+\tau^2)(m+kf)}\right)$ \\
\bottomrule
\end{tabular*}
\end{table*}

Using this notation, the main results of this article are provided in Table~1, which provides explicit expressions for Bayes factors based on $z$, $t$, $\chi^2$ and $F$ tests.  These expressions are justified by theorems provided in the Supplemental Material. 

In common applications of these tests, the convergence rates for the resulting Bayes factors may be summarized as follows. 
\begin{enumerate}
\item Under true alternative hypotheses (i.e., $\lambda \neq 0$), Bayes factors in favor of null hypotheses decrease exponentially fast to 0 as the sample size $n \rightarrow \infty$. 
\item Under true null hypotheses and certain regularity conditions, Bayes factors in favor of alternative hypotheses often decrease at rate $O_p(n^{-r-\half})$ for $z$ and $t$ tests and $O_p(n^{-r-\frac{k}{2}})$ for $\chi_k^2$ and $F_{k,m}$ tests. 
\end{enumerate}
Sufficient conditions for achieving these rates are provided in the Supplemental Material.

The improved convergence rates obtained for $r>1$ provide a partial motivation for generalizing the class of prior densities considered in J23. The non-local alternative prior densities proposed there correspond to setting $r=1$ in Table~1.  The improvement in convergence rates for true null hypotheses can be attributed to the more rapid descent of the non-local prior densities to 0 as the non-centrality parameters converge to $0$.

\section{Selection of prior hyperparameters}\label{sec3}
We now describe strategies for specifying $(\tau^2,r)$ to define either a Bayes factor or to construct a BFF. We assume that the non-centrality parameter $\lambda$ of the test statistic can be expressed as a function of $\omega$, a standardized effect. That is, $\lambda = \psi(\omega)$. For example, in a $z$ test that a normal mean $\mu$ is 0, the non-centrality parameter satisfies $\lambda =  \psi(\omega) = \sqrt{n}\omega$, where $\omega= \mu/\sigma$ and $n$ is the sample size.

Our recommendation for setting $(\tau^2,r)$ is to fix $\tau^2$ conditionally on $r$, and then set $\tau^2=\tau^2_{\omega,r}$ so that the prior mode on the non-centrality parameter equals $\psi(\omega^*)$, where $\omega^*$ represents a hypothesized value of $\omega$. That is, we define $\tau_{\omega^*,r}^2$ such that
\begin{equation}
    \psi(\omega^*) = \underset{\lambda}{\operatorname{argmax}}  \ \pi(\lambda|\tau^2_{\omega^*,r},r).
\end{equation}
Here, $\pi$ denotes the prior density on $\lambda$ under the alternative hypothesis. 
This constraint places the prior mode for $\lambda$ at the specified value of $\psi(\omega^*)$. In constructing a BFF, $\omega^*$ is varied over a plausible range of values for the standardized effect.

Like many default Bayesian testing methods, criteria for setting the scale of the prior density used to define alternative hypotheses remains a topic of active research \citep[e.g.,][]{Gprior_Zellner,Doucet2002,Liang08,Rouder2009,Consonni,Pramanik2023}. For the purposes of this article, we simply examine the sensitivity of Bayes factors and BFFs to the choice of $r$. Methods for estimating $r$ from published findings of similar studies or in replicated designs are currently under investigation.

To illustrate this strategy for setting $\tau^2$ for given $r$, consider
a $t$ test of a null hypothesis $H_0: \mu=0$ based on a random sample $x_1,\dots,x_n$, where $x_i \sim N(\mu,\sigma)$  and $\sigma^2$ is unknown. For this test, $t=\sqrt{n} \bar{x}/s$, where $s^2$ is the usual unbiased estimate of $\sigma^2$. The distribution of $t$ is
\begin{equation}\label{tex}
t\mid\mu, \sigma \sim T_{\nu}\left(\frac{\sqrt{n}\mu}{\sigma}\right),
\end{equation}
where $\nu=n-1$.
Under the null hypothesis, $t \sim T_{\nu}(0)$.  The non-centrality parameter of the $t$ distribution under the alternative hypothesis is $\lambda= \sqrt{n}\omega$, where we define $\omega=\mu/\sigma$ as the standardized effect.  

The prior distribution recommended for $\lambda$ in Table~1 is a normal-moment prior, $J(\tau^2,r)$. The modes of this density are $\pm\sqrt{2r}\tau$.  To define a Bayes factor given a value $\omega^*$, we select $\tau^2$ so that the modes of the prior occur at $\sqrt{n}\omega^*$.  That is, we equate $\sqrt{n}\omega^* = \sqrt{2r}\tau$ and define
$\tau^2_{\omega^*,r} = n{\omega^*}^2/(2r)$.

A similar procedure can be used to set $\tau^2_{\omega^*,r}$ for other test statistics. 
Table~\ref{tbl2} lists values of $\tau^2_{\omega^*,r}$ 
for several common statistical tests. This table generalizes the values provided in Table~1 of J23 for the case $r=1$.



\begin{figure}[ht!]
    \centering
    \includegraphics[width=8cm]{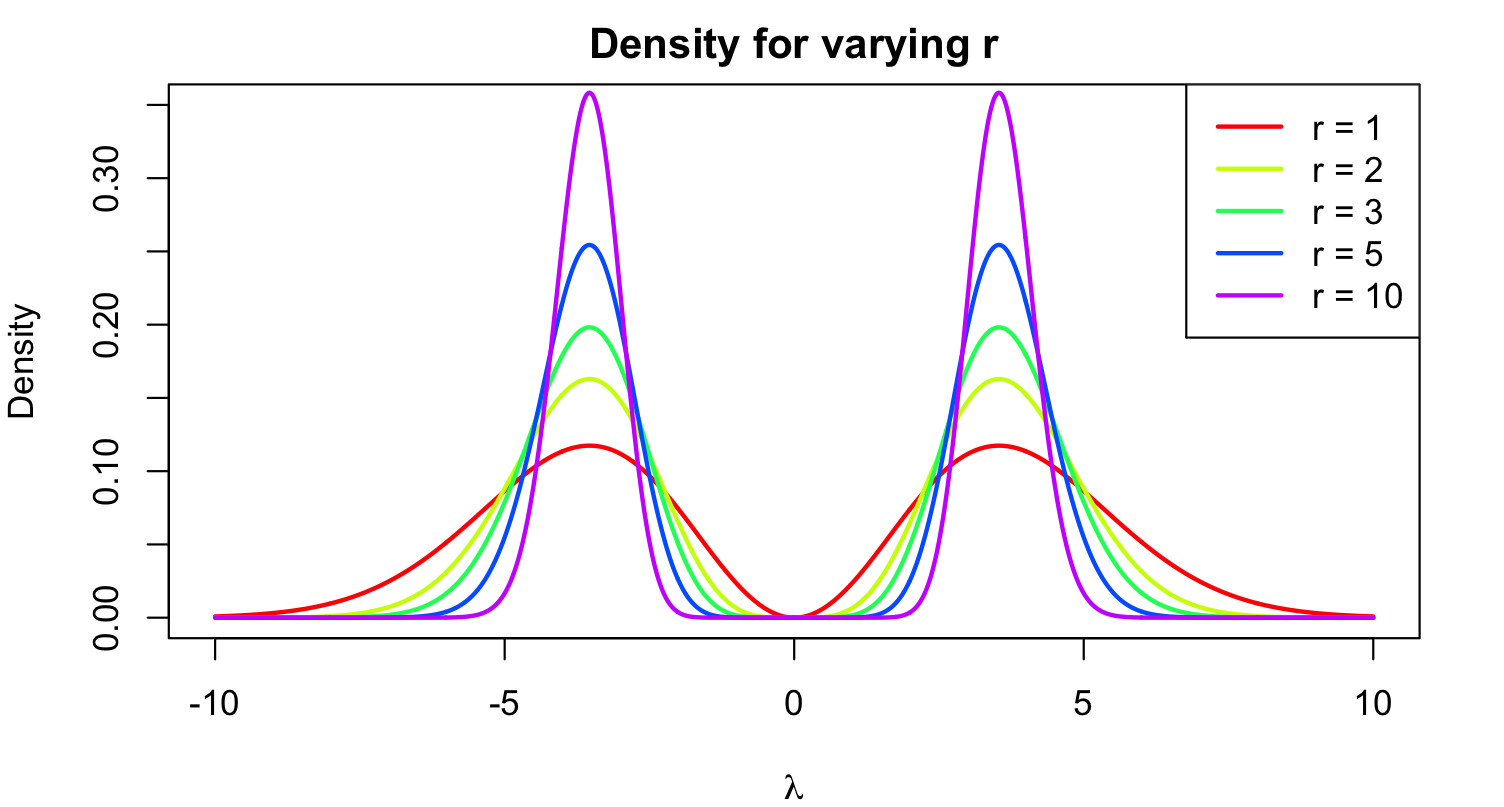}
    \caption{Normal moment prior densities with varying $r$}
    \label{nmr}
\end{figure}

Figure \ref{nmr} illustrates the effect of varying $r$ so that the mode of the prior density on an effect size remains constant.  It shows that the prior dispersion around the mode decreases as $r$ increases. 

\begin{table*}[width=1.9\linewidth,cols=4,pos=t]
\caption{Default choices of $\tau_{\omega,r}$}\label{tbl2}
\begin{tabular*}{\tblwidth}{@{} LCCC@{} }

\renewcommand{\arraystretch}{1.5} %
{\bf Test} & {\bf Statistic} & {\bf Standardized Effect ($\omega$)} & {\bf $\tau^2_{\omega,r}$} \\ \hline \hline
1-sample z & $\frac{\sqrt{n}\bar{x}}{\sigma}$ & $\frac{\mu}{\sigma}$ & $\frac{n\omega^2}{2r}$ \\  
1-sample t & $\frac{\sqrt{n}\bar{x}}{s}$ & $\frac{\mu}{\sigma}$ & $\frac{n\omega^2}{2r}$ \\  
2-sample z & $\frac{\sqrt{n_1 n_2}(\bar{x}_1-\bar{x}_2)}{\sigma\sqrt{n_1+n_2}}$ & $\frac{\mu_1-\mu_2}{\sigma}$ & $\frac{n_1 n_2\omega^2}{2r(n_1+n_2)}$ \\ 
2-sample t & $\frac{\sqrt{n_1 n_2}(\bar{x}_1-\bar{x}_2)}{s\sqrt{n_1+n_2}}$ & $\frac{\mu_1-\mu_2}{\sigma}$ & $\frac{n_1 n_2\omega^2}{2r(n_1+n_2)}$ \\ 
Multinomial/Poisson & $\chi^2_{\nu} = \sum\limits_{i=1}^k \frac{(n_i-nf_i(\hat{\theta}))^2}{nf_i(\hat{\theta})}$ & $\left(\frac{p_{i}-f_i(\theta)}{\sqrt{f_i(\theta)}}\right)_{k\times 1}$ & $\frac{n \omega'\omega}{2(\frac{k}{2}+r-1)} = \frac{nk\Tilde{\omega}^2}{2(\frac{k}{2}+r-1)}$ \\ 
Linear model & $F_{k,n-p} = \frac{(RSS_0-RSS_1)/k}{[(RSS_1)/(n-p)]}$ & $\frac{\mathbf{L}^{-1}(\mathbf{A}\boldsymbol{\beta}-\mathbf{a})}{\sigma}$ & $\frac{n \omega'\omega}{2(\frac{k}{2}+r-1)} = \frac{nk\Tilde{\omega}^2}{4(\frac{k}{2}+r-1)}$ \\ 
Likelihood Ratio & $\chi^2_{k} = -2\log\left[\frac{l(\theta_{r0},\hat{\theta_{s}})}{l(\hat{\theta})}\right]$ & ${\bf L}^{-1}(\theta_{r}-\theta_{r0})$ & $\frac{n \omega'\omega}{2(\frac{k}{2}+r-1)} = \frac{nk\Tilde{\omega}^2}{2(\frac{k}{2}+r-1)}$ \\  
\hline
\end{tabular*}
\end{table*}

\section{Examples}\label{sec:5}
To demonstrate the application of the Bayes factors described in Table~\ref{tbl1} and the procedure for setting $(\tau^2,r)$ described in Section~\ref{sec3}, we re-analyzed an experiment from the Many Labs~3 project \citep{ManyLabs3}. 
 
The experiment measured the effect on response time for subjects performing the Stroop task \citep{ManyLabs3}.  This effect is among the strongest and most widely replicated effects in experimental psychology. The Stroop task requires subjects to identify the color of the type of printed words. This task takes longer when there is discordance between the type's color and the color's name. For example, responding {\color{red} red} takes longer than {\color{red} blue}. After preprocessing the data to account for unusually long response times and incorrect answers \citep{Greenwald2003}, the authors ``calculated the average response time for all correct responses separately for congruent and incongruent trials" and then ``replaced response latencies for trials with errors using
the mean of correct responses in that condition plus 600 ms.'' They then ``recomputed the means for congruent and incongruent trials overall'' and used the 
difference between these two means divided by the standard deviation of all correct trials regardless of condition'' to construct paired $t$ statistics. 

Twenty replications of this experiment were replicated in the Many Labs experiment. For illustration, we begin by analyzing results from the first experimental site where the $t$ statistic was 9.38 on $83$ degrees of freedom (Table~3).

The null hypothesis in this study is that the mean difference in response times for the congruous and incongruous conditions is 0. Under the alternative hypothesis, we assume that the population mean difference in response times is $\mu$ and that the observational variance is $\sigma$, and let $\omega = \mu/\sigma$ denote the standardized effect size. Given $\omega >0$, we assume that the non-centrality parameters of the distributions of the $t$ statistic, $\lambda$, is drawn from a normal-moment distribution,
\begin{equation}\label{stroopprior}
  J^+\left(r+\frac{1}{2},\frac{1}{2\tau^2_{\omega,r}}\right) ,
\end{equation}
where, from Table~\ref{tbl2}, we set
\begin{equation}
    \tau^2_{\omega,r} = \frac{n \omega^2}{2r}.
\end{equation}

\begin{figure}[ht!]
    \centering
    \includegraphics[width=8cm]{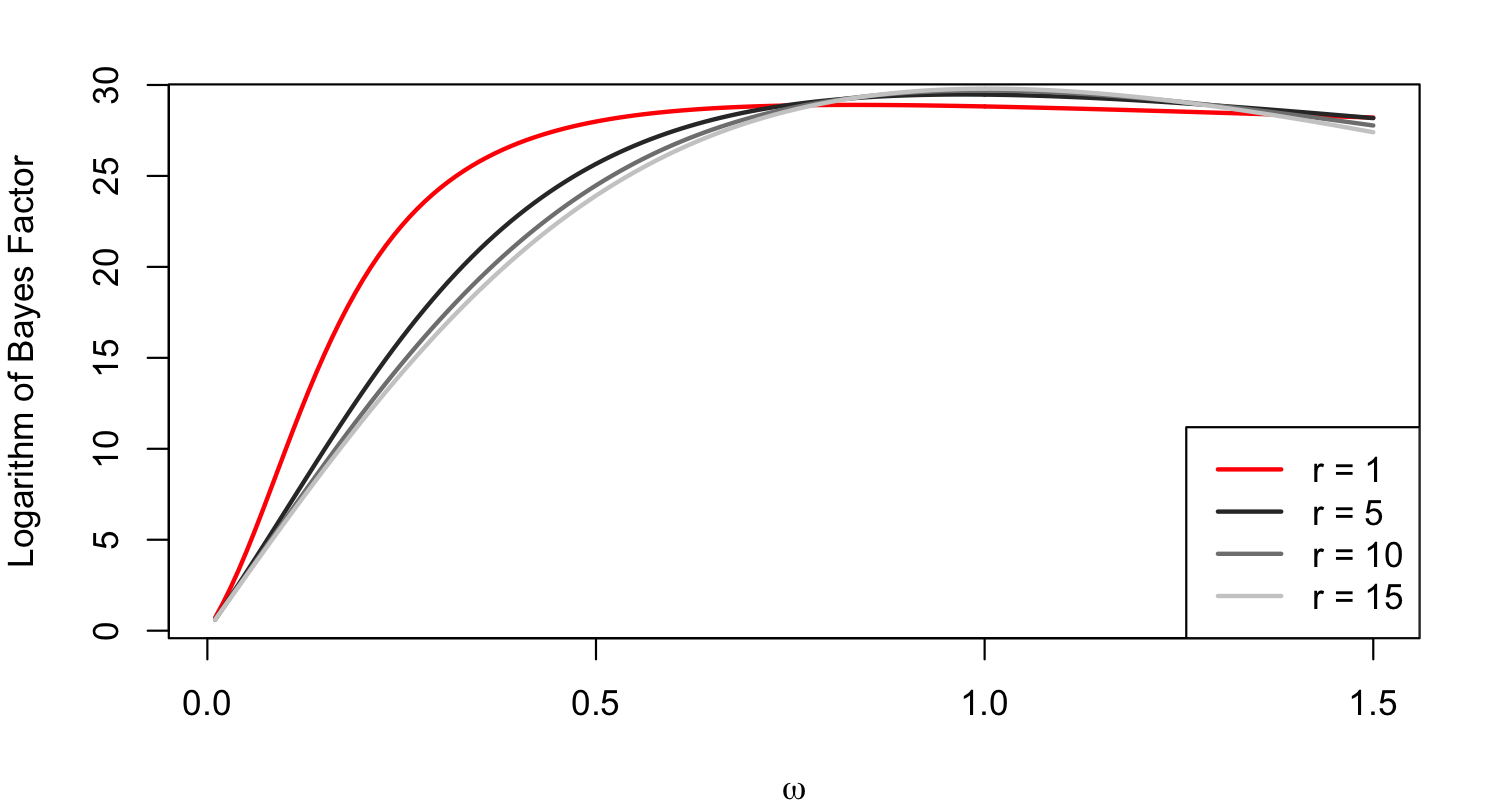}
    \caption{Bayes factor functions for Stroop test for varying $r$.}
    \label{StroopBFF}
\end{figure}

Figure \ref{StroopBFF} displays the plot of BFFs obtained for $r = 1, 5, 10$, and $15$ using the theoretical results from Tables~1 and~2.

Several aspects of this figure merit comment. As expected, the BFF curves reflect more evidence in favor of alternative hypotheses corresponding to $\omega \approx 1$ as $r$ increases. This happens because alternative hypotheses concentrate more of their prior mass on the hypothesized values of $\omega$ as $r$ increases. Indeed, as $r\rightarrow \infty$, all priors converge to a point mass on the hypothesized value of $\omega$, and the BFF curves reduce to a plot of the log-likelihood ratio (based on the test statistic). Nonetheless, all BFF curves reflect very strong evidence for alternative hypotheses defined by priors centered on standardized effect sizes greater than about 0.05, and the curves for $r=5,10,15$ are relatively insensitive to the choice of $r$ within this range.

We now consider how estimates of $r$ might be used to construct BFFs when data from replicated sites is available. In this case, data from 20 Many Labs consortium sites are available; $t$ statistics from the 20 sites are listed in Table~3.   For simplicity, we took a naive empirical Bayes approach to estimate $r$ (see Supplemental Materials), 
leading to an estimate of $r=9.99$. This value of $r$ suggests that standardized effects were relatively consistent across sites.  

Figure~3 provides the BFF based on the empirical Bayes estimate of $r$ and the $t$ statistics from all 20 sites.  For comparison, the BFF for $r=1$ is also displayed.  In this case, using the empirical Bayes estimate of $r$ had a moderate effect in increasing the BFF near its maximum of $\omega = 0.89$. With 20 experimental sites each exhibiting very strong evidence for a Stroop effect, the evidence against the null hypothesis is overwhelming except for alternative hypotheses concentrating prior mass on $\omega$ near 0.

\begin{table}
\centering
\caption{T statistics and degrees-of-freedom for the Stroop task [t-statistic (degrees of freedom)].}\label{ttable}

\begin{tabular}{|cccc|} \hline
  9.38(83) & 9.85(118) & 7.36(43) & 11.62(90)\\  7.85(95) & 12.56(317) & 11.01(123) & 10.15(130)\\  13.52(157) & 10.14(100) 
& 8.90(116) & 10.37(141) \\  11.68(177) & 9.11(118) & 16.97(241) & 8.82(136) \\  8.46(88) & 5.93(80) & 12.17(193) & 9.37(94) \\  
\hline
\end{tabular}

\end{table}

\begin{figure}[ht!]
    \centering
    \includegraphics[width=8cm]{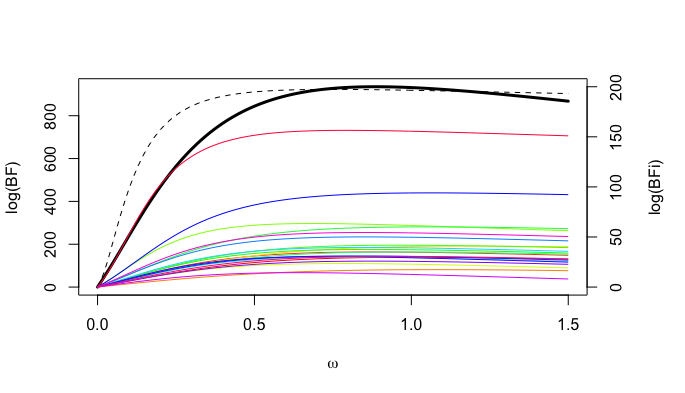}
    \caption{The combined BFF using the MOM estimator of \(r\) for the 20 sites in the test set is depicted by the solid black line. The combined BFF using \(r=1\) is shown by the black dashed line, while the BFFs for each individual study (using $r=1$) are illustrated by the colored lines.
}
    \label{BFF_Stroop}
\end{figure}

\section{Discussion}\label{sec:6}
Bayes factors based on test statistics present compelling advantages over standard methods for calculating Bayes factors. They eliminate computational challenges that arise when Bayes factors are computed from complex statistical models, and they reduce subjectivity when defining prior distributions on model parameters. By indexing Bayes factors based on test statistics according to standardized effects, BFFs eliminate much of the subjectivity associated with specifying a single alternative model. They also provide users with a simple representation of the statistical evidence in favor of alternative hypotheses centered on a range of effect sizes.

The BFFs proposed in J23 lack flexibility in allowing scientists to incorporate the precision of their estimates of effect sizes into Bayesian hypothesis tests. This article addresses this shortcoming by expanding the class of prior distributions used in defining BFFs. In particular, it provides analytic expressions for Bayes factors based on test statistics in conjunction with more general classes of prior distributions and illustrates how BFFs can be constructed from these broader classes. Importantly, we demonstrate how the scale parameters of the prior distributions can be linked so that the modes of prior distributions are located at hypothesized effect sizes.

J23 demonstrated how test statistics from replicated experiments could be combined to generate an aggregated BFF. This article's results will permit that methodology to account for the dispersion of effect sizes across replications of similar experiments. Efficient and coherent procedures for incorporating such information are currently under investigation.

The BFF package (\url{https://CRAN.R-project.org/package=BFF}), available from the CRAN depository, provides R functions to compute the BFFs reported in this article.

\section*{Acknowledgment}

The authors acknowledge support from the National Science Foundation, NSF DMS 2311005.



\bibliographystyle{plainnat}
\bibliography{Statistics_And_probability_letters/els-cas-templates/cas-refs}

\end{document}


\let\WriteBookmarks\relax
\def\floatpagepagefraction{1}
\def\textpagefraction{.001}
\shorttitle{}
\shortauthors{Datta, Shudde, Johnson}

\title [mode = title]{Supplemental Material for ``Bayes Factors Based on Test Statistics and Non-Local Moment Prior Densities''}                      

\tnotetext[1]{This document is the result of the research project funded by the National Science Foundation [DMS 2311005].}

\author[1]{Saptati Datta}[
                        orcid=0009-0009-3331-6523]

\credit{Conceptualization, Formal Analysis, Methodology, Software, Writing - original draft and revision}

\affiliation[1]{organization={Texas A\&M University},
                addressline={TAMU}, 
                postcode={77843}, 
                postcodesep={}, 
                city={College Station},
                country={USA}}

\author[1]{Rachael Shudde}[orcid=0000-0003-1672-3118]
\credit{Methodology, Software}

\author[1]{Valen E. Johnson}[orcid=0000-0002-8659-4772]
\cormark[1]
\ead{vejohnson@tamu.edu}

\credit{Conceptualization, Formal Analysis, Methodology, Software, Writing - original draft and revision, Supervision}



\begin{abstract}
    This supplement provides support for the Bayes factor expressions provided in Table 1 of "Bayes factors based on test statistics and non-local moment prior densities."  It also outlines the method-of-moment estimation procedure for the order of the normal moment priors.
\end{abstract}

\maketitle

\section*{Proofs of theorems supporting Table~1}
\textbf{Two-sided z-test:} Assume the distributions of a random variable $z$ under the null and alternative hypotheses are 
\begin{eqnarray}
H_0: z &\sim& N(0,1), \\
H_1: z| \lambda &\sim& N(\lambda,1), \ \lambda| \tau^2 \sim J(\tau^2, r), \ \tau>0, \ r \geq 1.
 \end{eqnarray}
 Then the Bayes factor in favor of the alternative hypothesis is
\begin{equation}
  BF_{10}(z|\tau^2) 
  = \frac{1}{(1+\tau^2)^{r + \frac{1}{2}}} {_1}F_1\Big(r+\frac{1}{2},\frac{1}{2}; \frac{\tau^2 z^2}{2(1+\tau^2)}\Big),
\end{equation} 
where ${_1}F_1(a,b;z)$ denotes the confluent hypergeometric function.

 \begin{proof} 
The density of $z$ under the null is given by 
\begin{equation}\label{eqn:50}
    m_0(z)= \frac{1}{\sqrt{2\pi}}\exp\big(-\frac{z^2}{2}\big)
\end{equation}
    Under the alternative, the density of $z \mid \lambda$ is,
    \begin{equation}
        p(z\mid\lambda) = \frac{1}{\sqrt{2\pi}}\exp\big[-\frac{(z-\lambda)^2}{2}\big]
    \end{equation}
    It follows that the marginal density under the alternative is 
 \begin{align}\label{eqn:51}
    m_1(z\mid \tau^2,r) &= \frac{1}{\sqrt{2\pi}(2\tau^2)^{r+\frac{1}{2}}\Gamma(r+\frac{1}{2})} \times \nonumber\\
    &\quad  \int_{-\infty}^{\infty} \left| \lambda \right|^{2r} \exp\left[-\frac{1}{2}\left\{(z-\lambda)^2+\frac{\lambda^2}{\tau^2}\right\}\right] \ d\lambda. \nonumber
\end{align}
    Let $a=\tau^{2} /\left(1+\tau^{2}\right)$, and note that

$$
\frac{\lambda^{2}}{\tau^{2}}+(z-\lambda)^{2}=\frac{1}{a}(\lambda-a z)^{2}+\frac{a z^{2}}{\tau^{2}}.
$$
 Using the expression for absolute fractional moments given in \cite{Fractional}, equation \eqref{eqn:51} leads  to 

\begin{align}
    m_1(z\mid \tau^2,r) &= \frac{\exp\big(-\frac{z^2}{2(1+\tau^2)}\big)}{\sqrt{2\pi}(2\tau^2)^{r+\frac{1}{2}}\Gamma(r+\frac{1}{2})} \times \nonumber \\
    &\quad  \int_{-\infty}^{\infty}\left| \lambda \right|^{2r} \exp\left(-\frac{1}{2a}(\lambda -az)^2\right) \ d\lambda \nonumber \\
    &= \frac{\exp\big(-\frac{z^2}{2(1+\tau^2)}\big)}{(2\tau^2)^{r+\frac{1}{2}}\Gamma(r+\frac{1}{2})\sqrt{\pi}} a^{r+\frac{1}{2}}\Gamma(r+\frac{1}{2}) \times \nonumber \\
    &\quad {_1}F_1\left(-r, \frac{1}{2}, -\frac{\tau^2 z^2}{2(1+\tau^2)}\right) \nonumber \\
    &= \frac{\exp\big(-\frac{z^2}{2(1+\tau^2)}\big)}{\sqrt{2\pi}(1+\tau^2)^{r+\frac{1}{2}}} \ {_1}F_1\left(-r,\frac{1}{2},-\frac{\tau^2 z^2}{2(1+\tau^2)}\right) \nonumber \\
    &= \frac{\exp\big(-\frac{z^2}{2}\big)}{\sqrt{2\pi}(1+\tau^2)^{r+\frac{1}{2}}} \ {_1}F_1\left(r+\frac{1}{2},\frac{1}{2},\frac{\tau^2 z^2}{2(1+\tau^2)}\right)
\end{align}

Dividing equation $m_1(z\mid \tau^2,r)$ by equation \eqref{eqn:50} yields the expression of the BFF for the two sided $z$ test.

\end{proof}
\emph{Rate of convergence for $z$-test:} 
Suppose that the following hold: 

\begin{description}
    \item[(i)] $z \sim N(\gamma \sqrt{n},1)$ for some $\gamma$ when $H_1$ is true ($\gamma>0$ in a one-sided test of $\mu>0$; $\gamma<0$ in a one-sided test of $\mu<0$),
    \item[(ii)]  $\tau^2 = \beta n $ for some $\beta>0$.
    \end{description}
    Then $BF_{01}(z \mid\tau^2, r)=O_p(\exp(-cn)) $ for some $c > 0$ when $H_1$ applies, and $BF_{10}(z \mid\tau^2, r) =O_p(n^{-r-\frac{ 1}{2}})$ when $H_0$ is true, $r$ being the true hyperparameter.
\begin{proof}
When the alternative is true,
\begin{eqnarray}
 BF_{10}(z \mid \tau^2, r)
 & = & \frac{1}{(1+\tau^2)^{r+\frac{1}{2}}}\sum_{i=0}^{\infty} \frac{(r+\frac{1}{2})^{(i)}}{(\frac{1}{2})^{(i)}i!} \Bigg(\frac{\tau^2 z^2}{2(1+\tau^2)}\Bigg)^i \nonumber \\
 & \geq & \frac{1}{(1+\tau^2)^{r+\frac{1}{2}}}\sum_{i=0}^{\infty} \frac{1}{i!} \Bigg(\frac{\tau^2 z^2}{2(1+\tau^2)}\Bigg)^i \nonumber \\
& = & \frac{1}{(1+\beta n)^{r+\frac{1}{2}}} \exp\Bigg(\frac{\tau^2 z^2}{2(1+\tau^2)}\Bigg) \nonumber 
\end{eqnarray}
This implies 
\begin{equation}
   BF_{01}(z \mid \tau^2, r) = O_p(\exp(-cn)), \quad \text{for some } c > 0 .
\end{equation}

    When the null is true,

    \begin{eqnarray}
  BF_{10}(z \mid \tau^2, r)
 & = & \frac{1}{(1+\beta n)^{r+\frac{1}{2}}}
{_1}F_1\Bigg(r+\frac{1}{2},\frac{1}{2}; \frac{\tau^2 z^2}{2(1+\tau^2)}\Bigg) \nonumber \\
& = & O_p(n^{-(r+\frac{1}{2})}).
    \end{eqnarray}

\end{proof}
\textbf{Two-sided t-test}:
     Assume the distributions of a random variable $t$ under the null and alternative hypotheses are 
\begin{eqnarray}
H_0: t &\sim& T_\nu(0), \\
H_1: t | \lambda &\sim&
 T_\nu(\lambda), \ \lambda | \tau^2 \sim J(\tau^2,r), \ \tau>0, \ r \geq 1.
 \end{eqnarray}
 Then the Bayes factor in favor of the alternative hypothesis is

\begin{multline}
    BF_{10}(t \mid \tau^2,r) = c \Bigg[  {_2}F_1\left(\frac{\nu+1}{2},r+\frac{1}{2},\frac{1}{2},y^2\right) \\
    + y\  \frac{\Gamma\left(\frac{\nu}{2}+1\right)}{\Gamma\left(\frac{\nu+1}{2}\right)}\frac{\Gamma(r+1)}{\Gamma\left(r+\frac{1}{2}\right)} \ \  {_2}F_1\left(\frac{\nu}{2}+1,r+1,\frac{3}{2},y^2\right) \Bigg]
\end{multline}
 where
 \begin{equation}\label{targs}
     y = \frac{\tau t}{\sqrt{(\nu+ t^2)(1+\tau^2})} \qquad \mbox{and} \qquad c = \frac{1}{(1+\tau^2)^{r+\frac{1}{2}}}.
 \end{equation}

Here ${_2}F_1(a,b,c; z)$ denotes the Gaussian hypergeometric function.  
\begin{proof}

When the alternative is true, the marginal distribution of the test statistic is

\begin{align}\label{eqn:0}
    m_1(t\mid \tau^2,r) 
    &= m_0(t)\Bigg[c\sum_{k=0}^\infty \frac{\Big(\frac{\nu+1}{2}\Big)^{(k)}}{(\frac{1}{2})^{(k)}k!}\Bigg(\frac{t^2}{2(t^2+\nu)}\Bigg)^k \times\nonumber\\
    &\quad  \int_{-\infty}^{\infty}\lambda^{2(r+k)}\exp\Big(-\frac{\lambda^2(\tau^2+1)}{2\tau^2}\Big) \, d\lambda + \nonumber \\ 
    &\quad  c\frac{\sqrt{2}t\Gamma\Big(\frac{\nu}{2}+1\Big)}{\sqrt{t^2+\nu}\Gamma\Big(\frac{\nu+1}{2}\Big)} \times \nonumber \\
    &\quad\sum_{k=0}^{\infty}\frac{\Big(\frac{\nu+1}{2}\Big)^{(k)}}{\big(\frac{3}{2}\big)^{(k)}k!}\Bigg(\frac{t^2}{2(t^2+\nu)}\Bigg)^k \times \nonumber \\
    &\quad  \int_{-\infty}^{\infty}\lambda^{2r+2k+1}\exp\Big(-\frac{\lambda^2(1+\tau^2)}{2\tau^2}\Big) \, d\lambda \Bigg]
\end{align}

    Let,
    \begin{eqnarray}
         I_1 
        & = & \int_{-\infty}^{\infty}\lambda^{2(r+k)}\exp\Big(-\frac{\lambda^2(\tau^2+1)}{2\tau^2}\Big) \ d\lambda \nonumber \\
        & = & 2\int_{0}^{\infty}\lambda^{2(r+k)}\exp\Big(-\frac{\lambda^2(\tau^2+1)}{2\tau^2}\Big) \ d\lambda \nonumber \\
        & = & \Gamma\Big(r+k+\frac{1}{2}\Big)\Big(\frac{2\tau^2}{1+\tau^2}\Big)^{r+k+\frac{1}{2}}
    \end{eqnarray}
    Assuming $\lambda^2=u$,
    \begin{eqnarray}
         I_2 
        & = &\int_{-\infty}^{\infty} \lambda^{2r+2k+1} \exp\Big(-\frac{\lambda^2(1+\tau^2)}{2\tau^2}\Big) \ d\lambda \nonumber \\ 
        & = & \frac{1}{2}\int_{0}^{\infty} u^{r+k} \exp\Big(-\frac{u(1+\tau^2)}{2\tau^2}\Big) \ du \nonumber \\
        & = &\frac{1}{2}\Gamma(r+k+1)\Big(\frac{2\tau^2}{1+\tau^2}\Big)^{r+k+1} \nonumber \\
        & = &\Gamma(r+k+1)2^{r+k}\Big(\frac{\tau^2}{1+\tau^2}\Big)^{r+k+1}
    \end{eqnarray}
    Substituting in \eqref{eqn:0}, the marginal density is
   \begin{align}
    &m_1(t\mid \tau^2,r) = m_0(t)\Bigg[ \frac{1}{(1+\tau^2)^{r+\frac{1}{2}}} \times \nonumber \\
    &\quad  {_2}F_1\left(\frac{v+1}{2},r+\frac{1}{2},\frac{1}{2},\frac{\tau^2 t^2}{(t^2+\nu)(1+\tau^2)}\right) + \nonumber \\
    &\quad  \frac{\tau t}{\sqrt{t^2+\nu}(1+\tau^2)^{r+1}}\frac{\Gamma\left(\frac{\nu}{2}+1\right)\Gamma(r+1)}{\Gamma\left(\frac{\nu+1}{2}\right)\Gamma\left(r+\frac{1}{2}\right)} \times \nonumber \\
    &\quad  {_2}F_1\left(\frac{\nu}{2}+1, r+1,\frac{3}{2},\frac{\tau^2 t^2}{(1+\tau^2)(t^2+\nu)}\right)\Bigg] \nonumber
\end{align}
    Hence the Bayes factor  based on two sided $t$ statistic is 
   \begin{align}
    &BF_{10}(t\mid \tau^2,r) = \frac{1}{(1+\tau^2)^{r+\frac{1}{2}}} \times \nonumber \\
    &\quad  {_2}F_1\left(\frac{v+1}{2},r+\frac{1}{2},\frac{1}{2},\frac{\tau^2 t^2}{(t^2+\nu)(1+\tau^2)}\right) + \nonumber \\
    &\quad \frac{\tau t}{\sqrt{t^2+\nu}(1+\tau^2)^{r+1}} \frac{\Gamma\left(\frac{\nu}{2}+1\right)\Gamma(r+1)}{\Gamma\left(\frac{\nu+1}{2}\right)\Gamma\left(r+\frac{1}{2}\right)} \times \nonumber \\
    &\quad  {_2}F_1\left(\frac{\nu}{2}+1, r+1,\frac{3}{2},\frac{\tau^2 t^2}{(1+\tau^2)(t^2+\nu)}\right).
\end{align}

\end{proof}
\emph{Rate of convergence for t-tests:}
Suppose the following apply: 
\begin{description}
    \item[(i)] $t \sim T_\nu(\gamma \sqrt{n})$ for some $\gamma$ when $H_1$ is true, ($\gamma>0$ in a one-sided test of $\mu>0$; $\gamma<0$ in a one-sided test of $\mu<0$),
    \item[(ii)] $\nu = \delta n - c$ for some $\delta,c > 0$, and 
      \item[(iii)] $\tau^2 = \beta n $ for some $\beta>0$.
    \end{description}
    
    Then $BF_{01}(t \mid \tau^2,r)=O_p(\exp(-cn)) $ for some $c > 0$ when the alternative hypothesis is true and $BF_{10}(t \mid \tau^2,r)= O_p(n^{-r-\frac{1}{2}})$ when the null hypothesis is true, $r$ being the true hyperparameter.
\begin{proof}
Re-write $BF_{10}(t\mid \tau^2, r)$ as 

\begin{align}
    &BF_{10}(t\mid \tau^2,r) = \frac{1}{(1+\tau^2)^{r+\frac{1}{2}}} \times \nonumber \\
    &\quad {_2}F_1\left(r+\frac{1}{2},\frac{v+1}{2},\frac{1}{2},\frac{\tau^2 t^2}{(t^2+\nu)(1+\tau^2)}\right) + \nonumber \\
    &\quad  \frac{\tau t}{\sqrt{t^2+\nu}\sqrt{1+\tau^2}(1+\tau^2)^{r+\frac{1}{2}}}  \frac{\Gamma\left(\frac{\nu}{2}+1\right)\Gamma(r+1)}{\Gamma\left(\frac{\nu+1}{2}\right)\Gamma\left(r+\frac{1}{2}\right)}\times \nonumber \\
    &\quad  {_2}F_1\left(r+1,\frac{\nu}{2}+1,\frac{3}{2},\frac{\tau^2 t^2}{(1+\tau^2)(t^2+\nu)}\right)
\end{align}

    It is known that  ${_1}F_1(a,c,z) =  \lim_{b\to\infty}  {_2}F_{1}(a,b,c,z/b)$.  Let \[ b_1 = \frac{\nu+1}{2}, \quad z_1 = \frac{(\nu+1)\tau^2 t^2}{2(1+\tau^2)(t^2+\nu)}, \quad a_1 = r+\frac{1}{2}, c_1 = \frac{1}{2}, \], 
    \[ b_2 = \frac{\nu}{2}+1,  z_2 = \frac{(\frac{\nu}{2}+1)\tau^2 t^2}{(1+\tau^2)(t^2+\nu)},  a_2=r+1, \ \mbox{and} \ c_2 = 3/2 .\] 
    
    When the alternative is true, $z_1, z_2$ are $O_p({n})$.
    As $b_1,b_2 \rightarrow \infty $, the Bayes factor in favour of the alternative can be approximated as 
  \begin{align}
    &BF_{10}(t \mid \tau^2,r) \approx \frac{1}{(1+\tau^2)^{r+\frac{1}{2}}} {_1}F_1(a_1, c_1, z_1) + \nonumber \\
    &\quad \frac{\tau t}{\sqrt{t^2+\nu} \sqrt{1+\tau^2} (1+\tau^2)^{r+\frac{1}{2}}} \nonumber \times  \\ &\quad \frac{\Gamma\left(\frac{\nu}{2}+1\right)\Gamma(r+1)}{\Gamma\left(\frac{\nu+1}{2}\right)\Gamma\left(r+\frac{1}{2}\right)}{_1}F_1(a_2, b_2, z_2) \nonumber \\
    &= \frac{1}{(1+\tau^2)^{r+\frac{1}{2}}} \sum_{i=0}^{\infty} \frac{\left(r+\frac{1}{2}\right)^{(i)}}{\left(\frac{1}{2}\right)^{(i)} i!} \left(\frac{(\nu+1)\tau^2 t^2}{2(1+\tau^2)(t^2+\nu)}\right)^i + \nonumber \\
    &\quad \frac{\tau t}{\sqrt{t^2+\nu} \sqrt{1+\tau^2} (1+\tau^2)^{r+\frac{1}{2}}} \times   \nonumber \\
    &\quad  \frac{\Gamma\left(\frac{\nu}{2}+1\right)\Gamma(r+1)}{\Gamma\left(\frac{\nu+1}{2}\right)\Gamma\left(r+\frac{1}{2}\right)} \sum_{i=0}^{\infty} \frac{(r+1)^{(i)}}{\left(\frac{3}{2}\right)^{(i)} i!} \left(\frac{\left(\frac{\nu}{2}+1\right)\tau^2 t^2}{(1+\tau^2)(t^2+\nu)}\right)^i \nonumber \\
    &\geq \frac{1}{(1+\tau^2)^{r+\frac{1}{2}}} \sum_{i=0}^{\infty} \left(\frac{(\nu+1)\tau^2 t^2}{2(1+\tau^2)(t^2+\nu)}\right)^i \frac{1}{i!} + \nonumber \\
    &\qquad \frac{\tau t}{\sqrt{t^2+\nu} \sqrt{1+\tau^2} (1+\tau^2)^{r+\frac{1}{2}}} \times\nonumber \\
    &\quad  \frac{\Gamma\left(\frac{\nu}{2}+1\right)\Gamma(r+1)}{\Gamma\left(\frac{\nu+1}{2}\right)\Gamma\left(r+\frac{1}{2}\right)}  \sum_{i=0}^{\infty} \left(\frac{\left(\frac{\nu}{2}+1\right)\tau^2 t^2}{(1+\tau^2)(t^2+\nu)}\right)^i \frac{1}{i!} .\nonumber
\end{align}

This implies 
\begin{align}\label{eqn:61}
    BF_{01}(t \mid \tau^2,r) 
    &= O_p(n^{r+\frac{1}{2}}) O_p(\exp(-cn))\nonumber \\ &\quad + O_p(n^{r+\frac{1}{2}}) O_p(\exp(-cn)) \nonumber \\
    &= O_p(\exp(-cn)), \quad \text{for some } c > 0 . \nonumber
\end{align}
    
    When the null is true, $z_1$ and $z_2$ are $O_p(1)$.
  \begin{align}
    BF_{10}(t \mid \tau^2, r) &\approx \frac{1}{(1+\tau^2)^{r+\frac{1}{2}}} {_1}F_1(a_1, c_1, z_1) + \nonumber \\
    &\quad \frac{\tau t}{\sqrt{t^2+\nu} \sqrt{1+\tau^2} (1+\tau^2)^{r+\frac{1}{2}}} \times \nonumber \\
    &\quad  \frac{\Gamma\left(\frac{\nu}{2}+1\right)\Gamma(r+1)}{\Gamma\left(\frac{\nu+1}{2}\right)\Gamma\left(r+\frac{1}{2}\right)} {_1}F_1(a_2, b_2, z_2) \nonumber \\
    &= \frac{1}{(1+\tau^2)^{r+\frac{1}{2}}} \times\nonumber \\ &\quad\sum_{i=0}^{\infty} \frac{\left(r+\frac{1}{2}\right)^{(i)}}{\left(\frac{1}{2}\right)^{(i)} i!} \left(\frac{(\nu+1)\tau^2 t^2}{2(1+\tau^2)(t^2+\nu)}\right)^i +  \nonumber \\
    &\quad  \frac{\tau t}{\sqrt{t^2+\nu} \sqrt{1+\tau^2} (1+\tau^2)^{r+\frac{1}{2}}} \times \nonumber \\
    &\quad  \frac{\Gamma\left(\frac{\nu}{2}+1\right)\Gamma(r+1)}{\Gamma\left(\frac{\nu+1}{2}\right)\Gamma\left(r+\frac{1}{2}\right)}\times \nonumber \\
    &\quad  \sum_{i=0}^\infty \frac{(r+1)^{(i)}}{\left(\frac{3}{2}\right)^{(i)} i!} \left(\frac{\left(\frac{\nu}{2}+1\right)\tau^2 t^2}{(1+\tau^2)(t^2+\nu)}\right)^i \nonumber \\
    &= O_p(n^{-r-\frac{1}{2}}) + O_p(n^{-r-\frac{1}{2}}) \nonumber \\
    &= O_p(n^{-r-\frac{1}{2}})
\end{align}

\end{proof}

\begin{itemize}
    \item 

\textbf{One-sided z-test:} Assume the distributions of a random variable $z$ under the null and alternative hypotheses are 
\begin{eqnarray}
H_0: z &\sim& N(0,1), \\
H_1: z| \lambda &\sim& N(\lambda,1), \ \lambda| \tau^2, r \sim J^+(\tau^2,r), \\ & & \ \ \tau>0, \ r\geq 1.
 \end{eqnarray}
 Then the Bayes factor in favor of the alternative hypothesis is

\begin{multline}\label{one_z}
  BF_{10}(z \mid \tau^2,r) = c \Bigg[ {_1}F_1\left(r+\frac{1}{2},\frac{1}{2},y^2\right) \\
  + 2 y \frac{\Gamma(r+1)}{\Gamma\left(r+\frac{1}{2}\right)} \ {_1}F_1\left(r+1, \frac{3}{2}, y^2\right) \Bigg]
\end{multline}
where
\begin{equation}
    y = \frac{\tau z}{\sqrt{2(1+\tau^2)}} \qquad \mbox{and} \qquad
    c = \frac{1}{(1+\tau^2)^{r+\frac{1}{2}}}.
\end{equation}
\item \textbf{One-sided t-test:}
     Assume the distributions of a random variable $t$ under the null and alternative hypotheses are 
\begin{eqnarray}
H_0: t &\sim& T(\nu,0), \\
H_1: t | \lambda &\sim&
 T(\nu,\lambda), \ \lambda | \tau^2 \sim J^{+}(2r,\tau^2), \\ & & \  \  \tau>0, \ r\geq 1.
 \end{eqnarray}
 Then the Bayes factor in favor of the alternative hypothesis is

\begin{equation}\label{one_t}
\begin{split}
    BF_{10}(t \mid \tau^2,r) &= c \Bigg[ {_2}F_1\left(\frac{\nu+1}{2},r+\frac{1}{2},\frac{1}{2},y^2\right) \\
    &\quad + 2y \frac{\Gamma\left(\frac{\nu}{2}+1\right)}{\Gamma\left(\frac{\nu+1}{2}\right)}\frac{\Gamma(r+1)}{\Gamma\left(r+\frac{1}{2}\right)} \\
    &\quad \times {_2}F_1\left(\frac{\nu}{2}+1,r+1,\frac{3}{2},y^2\right) \Bigg]
\end{split}
\end{equation}
 where $c$ and $y$ are defined in \eqref{targs}.
\end{itemize}
\begin{proof}\emph{One sided $z$ and $t$ tests:}
    
Under the null, the probability density function of $t$ is
$$m_0(t) = \frac{\Gamma\big(\frac{\nu + 1}{2}\big)}{\sqrt{\nu \pi}\Gamma\big(\frac{\nu}{2}\big)}\Big(1+\frac{t^2}{2}\Big)^{-\frac{\nu+1}{2}}.$$
The probability density function of a non-central t-distribution can be expressed using a power series expansion of the confluent hypergeometric function as
\begin{align}\label{62}
    m_1(t\mid \tau^2,r) &= 2 m_0(t) \Bigg[ c \sum_{k=0}^\infty \frac{\left(\frac{\nu+1}{2}\right)^{(k)}}{\left(\frac{1}{2}\right)^{(k)} k!} \left( \frac{t^2}{2(t^2+\nu)} \right)^k\times  \nonumber \\
    &\quad  \int_{0}^{\infty} \lambda^{2r+2k} \exp \left( -\frac{\lambda^2 (\tau^2 + 1)}{2\tau^2} \right) d\lambda + \nonumber \\
    &\quad c \frac{\sqrt{2}t \Gamma \left( \frac{\nu}{2} + 1 \right)}{\sqrt{t^2+\nu} \Gamma \left( \frac{\nu+1}{2} \right)} \times \nonumber \\ &\quad \sum_{k=0}^\infty \frac{\left( \frac{\nu+1}{2} \right)^{(k)}}{\left( \frac{3}{2} \right)^{(k)} k!} \left( \frac{t^2}{2(t^2+\nu)} \right)^k \times \nonumber \\
    &\quad \int_{0}^{\infty} \lambda^{2r+2k+1} \exp \left( -\frac{\lambda^2 (\tau^2 + 1)}{2\tau^2} \right) d\lambda \Bigg] \nonumber \\
    &= 2 m_0(t) (I_1 + I_2)
\end{align}

where $c = \frac{1}{(2\tau^2)^{r+\frac{1}{2}}\Gamma\big(r+\frac{1}{2}\big)}$.

Define
\begin{align}\label{63}
    I_1 &= c \sum_{k=0}^\infty \frac{\left(\frac{\nu+1}{2}\right)^{(k)}}{\left(\frac{1}{2}\right)^{(k)} k!} 
    \left(\frac{t^2}{2(t^2+\nu)}\right)^k \Gamma\left(r+k+\frac{1}{2}\right) \nonumber \\
    &\quad \times 2^{r+k-\frac{1}{2}} \left(\frac{\tau^2}{1+\tau^2}\right)^{r+k+\frac{1}{2}} \nonumber \\
    &= \left(\frac{\tau^2}{1+\tau^2}\right)^{r+\frac{1}{2}} \frac{1}{(2\tau^2)^{r+\frac{1}{2}}} 2^{r-\frac{1}{2}}\times \nonumber \\
    &\quad  \sum_{k=0}^\infty \frac{\left(\frac{\nu+1}{2}\right)^{(k)}}{\left(\frac{1}{2}\right)^{(k)} k!} 
    \left(\frac{t^2}{t^2+\nu}\right)^k \frac{\Gamma\left(r+k+\frac{1}{2}\right)}{\Gamma\left(r+\frac{1}{2}\right)} \nonumber \\
    &= \left(\frac{\tau^2}{1+\tau^2}\right)^{r+\frac{1}{2}} \frac{1}{(2\tau^2)^{r+\frac{1}{2}}} 2^{r-\frac{1}{2}} \times \nonumber \\
    &\quad  \sum_{k=0}^\infty \frac{\left(\frac{\nu+1}{2}\right)^{(k)} \left(r+\frac{1}{2}\right)^{(k)}}{\left(\frac{1}{2}\right)^{(k)} k!} 
    \left(\frac{t^2}{t^2+\nu}\right)^k \nonumber \\
    &= \frac{1}{2(1+\tau^2)^{r+\frac{1}{2}}} \times \nonumber \\ &\quad{_2}F_1\left(\frac{\nu+1}{2}, r+\frac{1}{2}, \frac{1}{2}, 
    \frac{\tau^2 t^2}{(t^2+\nu)(\tau^2+1)}\right)
\end{align}
and 
\begin{align}\label{64}
    I_2 &= c \frac{\sqrt{2}t \Gamma\left(\frac{\nu}{2}+1\right)}{\sqrt{t^2+\nu} \Gamma\left(\frac{\nu+1}{2}\right)} 
    \sum_{k=0}^\infty \frac{\left(\frac{\nu}{2}+1\right)^{(k)}}{\left(\frac{3}{2}\right)^{(k)} k!} 
    \left(\frac{t^2}{2(t^2+\nu)}\right)^k \times \nonumber \\
    &\quad 2^{r+k} \left(\frac{\tau^2}{1+\tau^2}\right)^{r+k+1} \Gamma(r+k+1) \nonumber \\
    &= \frac{1}{(2\tau^2)^{r+\frac{1}{2}} \Gamma\left(r+\frac{1}{2}\right)} 
    \frac{\sqrt{2}t \Gamma\left(\frac{\nu}{2}+1\right)}{\sqrt{t^2+\nu} \Gamma\left(\frac{\nu+1}{2}\right)} \times \nonumber \\ &\quad 2^r \left(\frac{\tau^2}{1+\tau^2}\right)^{r+1} \Gamma(r+1)  \times\nonumber \\
    &\quad  \sum_{k=0}^\infty \frac{\left(\frac{\nu}{2}+1\right)^{(k)}}{\left(\frac{3}{2}\right)^{(k)} k!} 
    \left(\frac{t^2}{t^2+\nu}\right)^k \left(\frac{\tau^2}{1+\tau^2}\right)^k 
    \frac{\Gamma(r+k+1)}{\Gamma(r+1)} \nonumber \\
    &= \frac{1}{(2\tau^2)^{r+\frac{1}{2}} \Gamma\left(r+\frac{1}{2}\right)} 
    \frac{\sqrt{2}t \Gamma\left(\frac{\nu}{2}+1\right)}{\sqrt{t^2+\nu} \Gamma\left(\frac{\nu+1}{2}\right)}  2^r \left(\frac{\tau^2}{1+\tau^2}\right)^{r+1} \times \nonumber \\
    &\quad \Gamma(r+1)  \sum_{k=0}^\infty \frac{\left(\frac{\nu}{2}+1\right)^{(k)} (r+1)^{(k)}}{\left(\frac{3}{2}\right)^{(k)} k!} 
    \left(\frac{t^2}{t^2+\nu}\right)^k \left(\frac{\tau^2}{1+\tau^2}\right)^k \nonumber \\
    &= \frac{t \tau}{\sqrt{t^2+\nu} (\tau^2+1)^{r+1}} 
    \frac{\Gamma\left(\frac{\nu}{2}+1\right)}{\Gamma\left(\frac{\nu+1}{2}\right)} 
    \frac{\Gamma(r+1)}{\Gamma\left(r+\frac{1}{2}\right)} \times \nonumber \\
    &\quad  {_2}F_1\left(\frac{\nu}{2}+1, r+1, \frac{3}{2}, 
    \frac{t^2 \tau^2}{(t^2+\nu)(1+\tau^2)}\right)
\end{align}
Combining equations \eqref{62}, \eqref{63} and \eqref{64}, the Bayes factor in favor of the alternative based on a one-sided t-test statistic is given by
\begin{align}
    &BF_{10}(t\mid \tau^2,r) = \frac{1}{(1+\tau^2)^{r+\frac{1}{2}}} \times \nonumber \\
    &\quad  {_2}F_1\left(\frac{\nu+1}{2}, r+\frac{1}{2}, \frac{1}{2}, \frac{\tau^2 t^2}{(t^2+\nu)(\tau^2+1)}\right) + \nonumber \\
    &\quad \frac{2t\tau}{\sqrt{t^2+\nu}(\tau^2+1)^{r+1}} \frac{\Gamma\left(\frac{\nu}{2}+1\right)}{\Gamma\left(\frac{\nu+1}{2}\right)} 
    \frac{\Gamma(r+1)}{\Gamma\left(r+\frac{1}{2}\right)} \times \nonumber \\
    &\quad {_2}F_1\left(\frac{\nu}{2}+1, r+1, \frac{3}{2}, \frac{t^2 \tau^2}{(t^2+\nu)(1+\tau^2)}\right).
\end{align}

Define, $w= \frac{(\nu+1)\tau^2 t^2}{(t^2+\nu)(1+\tau^2)}$. Then, $\frac{\tau^2 t^2}{(t^2+\nu)(\tau^2+1)}=\frac{2w}{\nu+1}.$
Note that $\lim_{b \rightarrow \infty}{_2}F_1(a,b,c,\frac{w}{b})={_1}F_1(a,c,w)$ and \newline ${_2}F_1(a,b,c,\frac{w}{b})={_2}F_1(b,a,c,\frac{w}{b})$ as $\nu \rightarrow \infty$, $w \rightarrow \frac{\tau^2 t^2}{2(1+\tau^2)}$. 

Hence, as $\nu \rightarrow \infty$,
\begin{eqnarray}
    & & \frac{1}{(1+\tau^2)^{r+\frac{1}{2}}}{_2}F_1\Bigg(\frac{\nu+1}{2},r+\frac{1}{2},\frac{1}{2},\frac{\tau^2 t^2}{(t^2+\nu)(\tau^2+1)}\Bigg) \nonumber \\ & = & \frac{1}{(1+\tau^2)^{r+\frac{1}{2}}}{_2}F_1\Bigg(r+\frac{1}{2},\frac{\nu+1}{2},\frac{1}{2},\frac{\tau^2 t^2}{(t^2+\nu)(\tau^2+1)}\Bigg) \nonumber \\ & \rightarrow &  \frac{1}{(1+\tau^2)^{r+\frac{1}{2}}}{_1}F_1\Bigg(r+\frac{1}{2},\frac{1}{2},\frac{\tau^2 t^2}{2(\tau^2+1)}\Bigg). \nonumber
\end{eqnarray}

    For large $\nu$, $\frac{\Gamma(\frac{\nu}{2}+1)}{\Gamma\big(\frac{\nu+1}{2}\big)}\approx \sqrt{\frac{\nu+1}{2}}$.  Hence, $\frac{\Gamma(\frac{\nu}{2}+1)}{\sqrt{t^2+\nu}\Gamma\big(\frac{\nu+1}{2}\big)}\approx \frac{\sqrt{\frac{\nu+1}{2}}}{\sqrt{t^2+\nu}}\rightarrow \frac{1}{\sqrt{2}} $. This implies 
    \begin{eqnarray}
      & \frac{2t\tau}{\sqrt{t^2+\nu}(\tau^2+1)^{r+1}}\frac{\Gamma(\frac{\nu}{2}+1)}{\Gamma\big(\frac{\nu+1}{2}\big)}\frac{\Gamma(r+1)}{\Gamma(r+\frac{1}{2})} {_2}F_1\Big(\frac{\nu}{2}+1,r+1,\frac{3}{2},\frac{t^2\tau^2}{(t^2+\nu)(1+\tau^2)}\Big) \nonumber \\
 &  \rightarrow \frac{2\tau t}{\sqrt{2}(\tau^2+1)^{r+1}}  \frac{\Gamma(r+1)}{\Gamma\big(r+\frac{1}{2}\big)} {_1}F_1\Big(r+1, \frac{3}{2}, \frac{\tau^2 t^2}{2(1+\tau^2)}\Big) \nonumber
    \end{eqnarray}
    Thus the Bayes factor function in favour of the alternative hypothesis based on a one-sided $z$-test statistic is 
 \begin{align}
    BF_{10}(z \mid \tau^2, r) 
    &= \frac{1}{(1+\tau^2)^{r+\frac{1}{2}}} {_1}F_1\left(r+\frac{1}{2}, \frac{1}{2}, \frac{\tau^2 z^2}{2(\tau^2+1)}\right) + \nonumber \\
    &\quad  \frac{2\tau z}{(\tau^2+1)^{r+1}} \frac{\Gamma(r+1)}{\Gamma\left(r+\frac{1}{2}\right)} \times \nonumber \\
    &\quad  {_1}F_1\left(r+1, \frac{3}{2}, \frac{\tau^2 z^2}{2(1+\tau^2)}\right).
\end{align}
\end{proof}
\textbf{$\chi^2$ test}:      Assume the distributions of a random variable $h$ under the null and alternative hypotheses are 
\begin{align}
H_0:  h &\sim \chi^2_k(0), \\
H_1: h \mid \lambda &\sim \chi^2_k(\lambda), \qquad
\lambda \mid \tau^2 \sim G\left( \frac{k}{2} + r, \frac{1}{2\tau^2} \right), \nonumber \\ \qquad
\tau &> 0, \ r \geq 1. \label{chisqprior}
\end{align}
Then the Bayes factor in favor of the alternative hypothesis is
\begin{equation}\label{x2bf}
    BF_{10}(h\mid \tau^2, r) = \frac{1}{(1+\tau^2)^{k/2+r}}  {_1}F_1\Big(\frac{k}{2}+r,\frac{k}{2};\frac{\tau^2 h}{2(1+\tau^2)}\Big) 
\end{equation}
\begin{proof}
    Under the null hypothesis, the density of $h$ is
\begin{equation}
\label{eqn:57}
m_0(h) = \frac{1}{2^\frac{k}{2} \Gamma(\frac{k}{2})} h^{\frac{k}{2} - 1} \exp(-\frac{h}{2}),
\end{equation}

and under the alternative hypothesis is
\begin{align}
    m_{1}(h\mid \tau^2,r) &= \int_{0}^{\infty} \sum_{i=0}^{\infty} 
    \frac{e^{-\lambda / 2}(\lambda / 2)^{i}}{i ! 2^{\frac{k}{2}+i} \Gamma\left(\frac{k}{2}+i\right)} 
    h^{\frac{k}{2}+i-1} e^{-h / 2} \times \nonumber \\
    &\quad  \frac{\lambda^{\frac{k}{2}+r-1} \exp \left(-\frac{\lambda}{2 \tau^{2}}\right)}
    {\left(2 \tau^{2}\right)^{\frac{k}{2}+r} \Gamma\left(\frac{k}{2}+r\right)} d \lambda.
\end{align}

This density can be simplified as

\begin{align}\label{eqn:58}
 m_1(h \mid \tau^2,r) 
&= \sum_{i=0}^{\infty} \frac{h^{\frac{k}{2}+i-1} \exp(-h / 2) 2^{-i}}{i ! 2^{\frac{k}{2}+i} \Gamma\left(\frac{k}{2}+i\right)\left(2 \tau^{2}\right)^{\frac{k}{2}+r} \Gamma\left(\frac{k}{2}+r\right)} \times  \nonumber \\
&\quad \int \lambda^{i+\frac{k}{2}+r-1} \exp \left(-\frac{\lambda}{2\left(1+\tau^{2}\right)}\right) d\lambda \nonumber \\
&= \sum_{i=0}^{\infty} \frac{h^{\frac{k}{2}+i-1} \exp(-h / 2) 2^{-i}}{i ! 2^{\frac{k}{2}+i} \Gamma\left(\frac{k}{2}+i\right)\left(2 \tau^{2}\right)^{\frac{k}{2}+r} \Gamma\left(\frac{k}{2}+r\right)}\times \nonumber \\
&\quad  \frac{\Gamma\left(\frac{k}{2}+r+i\right) 2^{\frac{k}{2}+r+i}}{\left(1+1 / \tau^{2}\right)^{\frac{k}{2}+r+i}} \nonumber \\
&= \frac{h^{\frac{k}{2}-1} \exp(-h / 2)}{2^{\frac{k}{2}} \Gamma\left(\frac{k}{2}+r\right)\left(1+\tau^{2}\right)^{\frac{k}{2}+r}} \times \nonumber \\&\quad \sum_{i=0}^{\infty} \left[\frac{\tau^{2} h}{2\left(1+\tau^{2}\right)}\right]^{i} \frac{1}{i !} \times\nonumber \\
&\qquad  \left(\frac{k}{2}+r+i-1\right) \cdots \left(\frac{k}{2}+i\right) \nonumber \\
&= \frac{h^{\frac{k}{2}-1} \exp(-h / 2)}{2^{\frac{k}{2}} \Gamma\left(\frac{k}{2}+r\right)\left(1+\tau^{2}\right)^{\frac{k}{2}+r}} \times \nonumber \\ &\qquad \sum_{i=0}^{\infty} \left[\frac{\tau^{2} h}{2\left(1+\tau^{2}\right)}\right]^{i} \frac{1}{i!} \times \nonumber \\
&\qquad  \frac{\Gamma\left(\frac{k}{2}+r+i\right)}{\Gamma\left(\frac{k}{2}+i\right)} \nonumber \\
&= \frac{h^{\frac{k}{2}-1} \exp(-h / 2)}{2^{\frac{k}{2}} \Gamma\left(\frac{k}{2}\right)\left(1+\tau^{2}\right)^{\frac{k}{2}+r}} \nonumber \\ &\qquad \times \sum_{i=0}^\infty \frac{\left(\frac{k}{2}+r\right)^{(i)}}{\left(\frac{k}{2}\right)^{(i)}} \left[\frac{\tau^{2} h}{2\left(1+\tau^{2}\right)}\right]^{i} \frac{1}{i!} \nonumber \\
&= \frac{h^{\frac{k}{2}-1} \exp(-h / 2)}{2^{\frac{k}{2}} \Gamma\left(\frac{k}{2}\right)\left(1+\tau^{2}\right)^{\frac{k}{2}+r}} \nonumber\\  &\qquad \times {_1}F_1 \left(\frac{k}{2}+r, \frac{k}{2}, \frac{\tau^2 h }{2(1+\tau^2)}\right).
\end{align}
Dividing equation \eqref{eqn:58} by equation \eqref{eqn:57}, we get the expression in BFBOTS for a $\chi^2$ test.
\end{proof}
\emph{Rate of Convergence for $\chi^2$ test:}
Suppose the following hold: 
\begin{description}
    \item[(i)] $h \sim \chi^2_k (\gamma n)$ for some $\gamma>0$ when the alternative hypothesis is true, and
       \item[(iii)]  $\tau^2 = \beta n $ for some $\beta>0$.
    \end{description}
    Then $BF_{01}(h \mid \tau^2,r)=O_p[\exp(-cn)] $ for some $c > 0$ when the alternative hypothesis is true and $BF_{10}(h \mid \tau^2,r)=O_p(n^{-r-\frac{k}{2}})$ when the null hypothesis is true, $r$ being the true hyperparameter.

\begin{proof}
  When the alternative is true,
\begin{eqnarray}
 BF_{10}(h \mid \tau^2, r)
 & = & \frac{1}{(1+\tau^2)^{r+\frac{k}{2}}}\sum_{i=0}^{\infty} \frac{(r+\frac{k}{2})^{(i)}}{(\frac{k}{2})^{(i)}i!} \Bigg(\frac{\tau^2 h}{2(1+\tau^2)}\Bigg)^i \nonumber \\
 & \geq & \frac{1}{(1+\tau^2)^{r+\frac{k}{2}}}\sum_{i=0}^{\infty} \frac{1}{i!} \Bigg(\frac{\tau^2 h}{2(1+\tau^2)}\Bigg)^i \nonumber \\
& = & \frac{1}{(1+\beta n)^{r+\frac{k}{2}}} \exp\Bigg(\frac{\tau^2 h}{2(1+\tau^2)}\Bigg) \nonumber \\
\end{eqnarray}
Thus,
\[ 
BF_{01}(h \mid \tau^2, r) = O_p(\exp(-cn)), \quad {c > 0}
\]

    When the null is true,

    \begin{eqnarray}
  BF_{10}(h \mid \tau^2, r)
 & = & \frac{1}{(1+\beta n)^{r+\frac{k}{2}}}
{_1}F_1\Bigg(r+\frac{k}{2},\frac{k}{2}; \frac{\tau^2 z^2}{2(1+\tau^2)}\Bigg) \nonumber \\
& = & O_p(n^{-(r+\frac{k}{2})})
   \end{eqnarray}
    \end{proof}

\textbf{F test:} Assume the distributions of a random variable $f$ under the null and alternative hypotheses are 
\begin{align}
H_0:  f &\sim F_{k,m}(0), \\
H_1: f \mid \lambda &\sim F_{k,m}(\lambda),\lambda \mid \tau^2 \sim G\left( \frac{k}{2} + r, \frac{1}{2\tau^2} \right), \nonumber \\
&\tau > 0, \ r \geq 1. \label{falt}
\end{align}

Then the Bayes factor in favor of the alternative hypothesis is
\begin{multline}\label{eqn:f}
    BF_{10}(f \mid \tau^2, r) =  \frac{1}{(1+\tau^2)^{\frac{k}{2}+r}} \\
    \times {_2}F_1\left(\frac{k}{2}+r, \frac{k+m}{2}, \frac{k}{2}; \frac{kf\tau^2}{(1+\tau^2)(m+kf)}\right)
\end{multline}

\begin{proof}
Under the null hypothesis, the density of $f$ is

\begin{equation}
\label{eqn:6}
    m_0(f)=\frac{1}{\beta(\frac{k}{2},\frac{m}{2})}\Big(\frac{k}{m}\Big)^{\frac{k}{2}} f^{\frac{k}{2} - 1}\Big(1+\frac{k}{m}f\Big)^{-\frac{k+m}{2}}.
\end{equation}

Under the alternative hypothesis, the marginal density of $f$ is

\begin{align}\label{eqn:60} 
    m_1(f\mid \tau^2,r) 
    &= \int_{0}^{\infty} \sum_{i=0}^{\infty} 
    \frac{\exp \left[-\frac{\lambda}{2}\right] \left(\frac{\lambda}{2}\right)^{i} \Gamma\left(\frac{k}{2}+\frac{m}{2}+i\right)}
    {i! \Gamma\left(\frac{k}{2}+i\right) \Gamma\left(\frac{m}{2}\right)} \times \nonumber \\
    &\quad  \left(\frac{k}{m}\right)^{\frac{k}{2}+i}\left(\frac{m}{m+kf}\right)^{\frac{k}{2}+\frac{m}{2}+i} f^{\frac{k}{2}+i-1} \times\nonumber \\
    &\quad  \frac{\lambda^{\frac{k}{2}+r-1} \exp\left[-\frac{\lambda}{2\tau^2}\right]}
    {\Gamma\left(\frac{k}{2}+r\right) \left(2\tau^2\right)^{\frac{k}{2}+r}} \, d\lambda \nonumber \\
    &= \sum_{i=0}^{\infty} 
    \frac{\Gamma\left(\frac{k}{2}+\frac{m}{2}+i\right)}{2^i i! \Gamma\left(\frac{k}{2}+i\right) \Gamma\left(\frac{m}{2}\right)} 
    \left(\frac{k}{m}\right)^{\frac{k}{2}+i} \times\nonumber \\
    &\quad  f^{\frac{k}{2}+i-1} \left(\frac{m}{m+kf}\right)^{\frac{k}{2}+\frac{m}{2}+i} \times\nonumber \\
    &\quad \frac{\Gamma\left(i+\frac{k}{2}+r\right)}
    {\Gamma\left(\frac{k}{2}+r\right) \left(2\tau^2\right)^{\frac{k}{2}+r}} \left(\frac{2\tau^2}{\tau^2 + 1}\right)^{i+\frac{k}{2}+r} \nonumber \\
    &= \frac{\Gamma\left(\frac{k+m}{2}\right)}{\Gamma\left(\frac{k}{2}\right) \Gamma\left(\frac{m}{2}\right)} 
    \left(\frac{k}{m}\right)^{\frac{k}{2}} \left(1+\frac{kf}{m}\right)^{-\frac{k+m}{2}} \times \nonumber \\
    &\quad  \left(\frac{1}{1+\tau^2}\right)^{\frac{k}{2}+r} f^{\frac{k}{2}-1} \times \nonumber \\ &\quad  \sum_{i=0}^{\infty} 
    \frac{\left(\frac{k+m}{2}\right)^{(i)} \left(\frac{k}{2}+r\right)^{(i)}}{\left(\frac{k}{2}\right)^{(i)} i!} 
    \left(\frac{kf\tau^2}{(m+kf)(1+\tau^2)}\right)^i \nonumber \\
    &= \frac{m_0(f)}{(1+\tau^2)^{\frac{k}{2}+r}} \times \nonumber \\ &\qquad 
    {_2}F_{1}\left(\frac{k+m}{2}, \frac{k}{2}+r, \frac{k}{2}, \frac{kf\tau^2}{(m+kf)(1+\tau^2)}\right)
\end{align}

Dividing the resulting expression in equation \eqref{eqn:60} by $m_0(f)$, we get the expression of the Bayes factor based on an $F$-statistic.
 \end{proof}  

 \emph{Rate of Convergence for $F$-test}
Suppose the following hold: 
\begin{description}
    \item[(i)] $f \sim F_{k,m}(\gamma n)$ for some $\gamma>0$ when the alternative hypothesis is true, and
    \item[(ii)] $m = \delta n - c$ for some $\delta,c > 0$.
        \item[(iii)]  $\tau^2 = \beta n $ for some $\beta>0$.
    \end{description}
    
    Then $BF_{01}(f \mid\tau^2,r)=O_p[\exp(-n)]$  when the alternative hypothesis is true and $BF_{10}(f \mid\tau^2,r) = O_p(n^{-1-\frac{k}{2}})$ when the null hypothesis is true, $r$ being the true hyperparameter.
\begin{proof}
 Define $w = \frac{(k+m)kf\tau^2}{2(m+kf)(1+\tau^2)}$. When the alternative is true, \newline
 $w = O_p(cn),c>0$.  This implies
 \begin{align}
    BF_{10}(f \mid \tau^2,r) 
    &= \frac{1}{(1+\tau^2)^{\frac{k}{2}+r}} \times\nonumber \\ &\quad
    {_2}F_{1}\left(\frac{k+m}{2}, \frac{k}{2}+r, \frac{k}{2}, \frac{kf\tau^2}{(m+kf)(1+\tau^2)}\right) \nonumber \\
    &= \frac{1}{(1+\tau^2)^{\frac{k}{2}+r}} \times\nonumber \\ &\quad 
    {_2}F_{1}\left(\frac{k}{2}+r, \frac{k+m}{2}, \frac{k}{2}, \frac{kf\tau^2}{(m+kf)(1+\tau^2)}\right) \nonumber \\
    &\approx \frac{1}{(1+\tau^2)^{\frac{k}{2}+r}} 
    {_1}F_{1}\left(\frac{k}{2}+r, \frac{k}{2}, z\right) \nonumber \\
    &= \frac{1}{(1+\tau^2)^{\frac{k}{2}+r}} \times\nonumber \\ &\quad 
    \sum_{i=0}^{\infty} \frac{\left(\frac{k}{2}+r\right)^{(i)}}{\left(\frac{k}{2}\right)^{(i)} i!} 
    \left(\frac{(k+m)kf\tau^2}{2(m+kf)(1+\tau^2)}\right)^i \nonumber \\
    &\geq \frac{1}{(1+\tau^2)^{\frac{k}{2}+r}} 
    \sum_{i=0}^{\infty} \left(\frac{(k+m)kf\tau^2}{2(m+kf)(1+\tau^2)}\right)^i 
    \frac{1}{i!} \nonumber
\end{align}

    Thus,
    \begin{eqnarray} 
    BF_{01}(f \mid \tau^2,r) & = & O(n^{\frac{k}{2}+r}) O_p(\exp(-cn)), c>0 \nonumber \\
    & = & O_p(\exp(-cn)) \nonumber
 \end{eqnarray}
     When the null is true, $w = O_p(1)$.
  \begin{align}
    BF_{10}(f \mid \tau^2,r) 
    &= \frac{1}{(1+\tau^2)^{\frac{k}{2}+r}} \times\nonumber \\ &\quad 
    {_2}F_{1}\left(\frac{k+m}{2}, \frac{k}{2}+r, \frac{k}{2}, \frac{kf\tau^2}{(m+kf)(1+\tau^2)}\right) \nonumber \\
    &= \frac{1}{(1+\tau^2)^{\frac{k}{2}+r}} \times \nonumber \\
    &\quad  {_2}F_{1}\left(\frac{k}{2}+r, \frac{k+m}{2}, \frac{k}{2}, \frac{kf\tau^2}{(m+kf)(1+\tau^2)}\right) \nonumber \\
    &\approx \frac{1}{(1+\tau^2)^{\frac{k}{2}+r}} {_1}F_{1}\left(\frac{k}{2}+r, \frac{k}{2}, w\right) \nonumber \\
    &= \frac{1}{(1+\tau^2)^{\frac{k}{2}+r}} \times\nonumber \\
    &\quad  \sum_{i=0}^{\infty} \frac{\left(\frac{k}{2}+r\right)^{(i)}}{\left(\frac{k}{2}\right)^{(i)} i!} 
    \left(\frac{(k+m)kf\tau^2}{2(m+kf)(1+\tau^2)}\right)^i \nonumber \\
    &= O(n^{-\frac{k}{2}+r}) O_p(1) \nonumber \\
    &= O_p(n^{-\frac{k}{2}+r}) .\nonumber
\end{align}
 \end{proof}

\section*{Empiricial Bayes estimation of $r$}

We describe Method of Moments (MOM) estimators for \( r \) that are independent of standardized effect sizes. Since \( r \) represents a measure of variability across studies, it cannot be estimated from a single study. When dealing with a single study, \( r \) should be chosen using prior knowledge of the variability of effect sizes around their population value.

\medskip

\emph{Choice of $r$ for a Z-test.}
Under $H_1$, $Z_i \mid \lambda_i \sim N(\lambda_i, 1)$ and $\lambda_i \sim J\left(r, \frac{1}{2\tau_i^2}\right)$, $i=1,2,\ldots, S$. Define, $\theta_i = \lambda_i^2$.

Hence, $X_i = Z_i^2 \mid \theta_i \sim \chi^2_1(\theta_i)$ and $\theta_i \sim G\left(r+\frac{1}{2}, \frac{1}{2\tau_i^2}\right)$.

\begin{equation*}
    E(X_i) = E[E(X_i \mid \theta_i)] = \left(r+\frac{1}{2}\right)2\tau_i^2 = \left(r+\frac{1}{2}\right)\frac{n_i \omega^2}{r}.
\end{equation*}
This implies,
\begin{equation}
    E\left(\frac{X_i}{n_i}\right) = \left(r+\frac{1}{2}\right)\frac{\omega^2}{r}.
\end{equation}
Now,
 \begin{eqnarray}
     Var(X_i) & = & Var(E(X_i \mid \theta_i)) + E(Var(X_i \mid \theta_i)) \nonumber \\
     & = & Var(1+\theta_i) + E(Var(1+2\theta_i)) \nonumber \\
     & = & \left(r+\frac{1}{2}\right)(2\tau_i^2)^2 + 2 + 4\left(r + \frac{1}{2}\right)2\tau_i^2 
     \end{eqnarray}
Hence,
\begin{equation}
    Var\left(\frac{X_i}{n_i}\right) = \left(r+\frac{1}{2}\right)\frac{\omega^4}{r^2} + \frac{2}{n_i^2} + 4\left(r + \frac{1}{2}\right)\frac{\omega^2}{n_i} 
\end{equation}

Ignoring the $O\left(\frac{1}{n_i}\right)$ terms, the MOM estimator of $r$ is,
\begin{equation}
    \hat{r}_{MOM} =\left[ \hat{E}^2\left(\frac{x_i}{n_i}\right)/\hat{V}\left(\frac{x_i}{n_i}\right)\right] - \frac{1}{2},
\end{equation}
where $z_i'$s are the observed test statistics, $x_i = z_i^2$, $i = 1,2,\ldots,S$, \[\hat{E}\left(\frac{x_i}{n_i}\right) = \frac{1}{S} \sum_{i=1}^S \left(\frac{x_i}{n_i}\right)\] and \[\hat{V}\left(\frac{x_i}{n_i}\right) = \frac{1}{S}\sum_{i=1}^S \left(\frac{x_i}{n_i} - \frac{1}{S} \sum_{i=1}^S \left(\frac{x_i}{n_i}\right)\right)^2.\]

     \emph{Choice of $r$ for a t-test.}
     Under $H_1$, $t_i \mid \lambda_i \sim T_{\nu_i}(\lambda_i)$ and $\lambda_i \sim J\left(r, \frac{1}{2\tau_i^2}\right)$, $i=1,2,\ldots, S$. Define, $\theta_i = \lambda_i^2$.

Hence, $X_i = t_i^2 \mid \theta_i \sim F_{1,\nu_i}(\theta_i)$ and $\theta_i \sim G\left(r+\frac{1}{2}, \frac{1}{2\tau_i^2}\right)$.

\begin{equation*}
    E(X_i) = E[E(X_i \mid \theta_i)] = \frac{\nu_i}{\nu_i-2}\left(1+\left(r+\frac{1}{2}\right)2\tau_i^2 \right).
\end{equation*}
This implies,
\begin{equation}
    E\left(\frac{X_i}{n_i}\right) = \frac{\nu_i}{\nu_i-2}\left(\frac{1}{n_i}+\left(r+\frac{1}{2}\right)\frac{ \omega^2}{r}\right)
\end{equation}
Now,
\begin{eqnarray}
    Var(X_i) & = & Var(E(X_i \mid \theta_i)) + E(Var(X_i \mid \theta_i)) \nonumber \\
    & = & Var\left(\frac{\nu_i(1+\theta_i)}{\nu_i-2}\right) \nonumber \\
    & & + E\left(2\nu_i^2\frac{(1+\theta_i)^2 + (1+2\theta_i)(\nu_i-2)}{(\nu_i-2)^2(\nu_i-4)}\right) \nonumber \\
    & = & \left(\frac{\nu_i}{\nu_i-2}\right)^2 \left(r+\frac{1}{2}\right)(2\tau_i^2)^2 \nonumber \\
    & & + 2 \left(\frac{\nu_i}{\nu_i-2}\right)^2\frac{1}{\nu_i-4} \left[ 1 + \left(r+\frac{1}{2}\right)(2\tau_i^2)^2 \right. \nonumber \\
    & & \left. + \left(r+\frac{1}{2}\right)^2(2\tau_i^2)^2 + (\nu_i - 2) + 2(\nu_i-2)2\tau_i^2 \right] \nonumber
\end{eqnarray}
Hence, assuming $\nu_i = O(n_i)$,
\begin{equation}
    Var\left(\frac{X_i}{n_i}\right) = \left(\frac{\nu_i}{\nu_i-2}\right)^2 \left(r+\frac{1}{2}\right)\frac{\omega^4}{r^2} + O\left(\frac{1}{n_i}\right).
\end{equation}

Ignoring the $O\left(\frac{1}{n_i}\right)$ terms, the MOM estimator of $r$ is,
\begin{equation}
    \hat{r}_{MOM} =\left[ \hat{E}^2\left(\frac{x_i}{n_i}\right)/\hat{V}\left(\frac{x_i}{n_i}\right)\right] - \frac{1}{2},
\end{equation}
where $t_i'$s are the observed test statistics, $x_i = t_i^2$, $i = 1,2,\ldots,S$, \[\hat{E}\left(\frac{x_i}{n_i}\right) = \frac{1}{S} \sum_{i=1}^S \left(\frac{x_i}{n_i}\right)\] and \[\hat{V}\left(\frac{x_i}{n_i}\right) = \frac{1}{S}\sum_{i=1}^S \left(\frac{x_i}{n_i} - \frac{1}{S} \sum_{i=1}^S \left(\frac{x_i}{n_i}\right)\right)^2.\]

\emph{Choice of $r$ for a $\chi^2$ test.}
 Under $H_1$, $H_i \mid \lambda_i \sim \chi^2_{k}(\lambda_i)$ and $\lambda_i \sim G\left(\frac{k}{2}+r, \frac{1}{2\tau_i^2}\right)$, $i=1,2,\ldots, S$.

\begin{equation*}
    E(H_i) = E[E(H_i \mid \lambda_i)] = k+\left(\frac{k}{2}+r\right)2\tau_i^2.
\end{equation*}
This implies,
\begin{equation}
    E\left(\frac{H_i}{n_i}\right) = \frac{k}{n_i}+\left(\frac{k}{2}+r\right)\frac{\Tilde{\omega}^2}{r}.
\end{equation}
Now,
 \begin{eqnarray}
     Var(H_i) & = & Var(E(H_i \mid \lambda_i)) + E(Var(H_i \mid \lambda_i)) \nonumber \\
     & = & \left(\frac{k}{2}+r\right)(2\tau_i^2)^2 + 2k + 4\left(\frac{k}{2}+r\right)(2\tau_i^2)
     \end{eqnarray}
Hence, assuming $k = O(1)$,
\begin{equation}
    Var\left(\frac{H_i}{n_i}\right) =  \left(\frac{k}{2}+r\right)\frac{\omega^4}{r^2} + 2\frac{k}{n_i} + 4\left(\frac{k}{2}+r\right)\left(\frac{\Tilde{\omega}^2}{n_i r}\right)
\end{equation}

Ignoring the $O\left(\frac{1}{n_i}\right)$ terms, the MOM estimator of $r$ is,
\begin{equation}
    \hat{r}_{MOM} =\left[ \hat{E}^2\left(\frac{h_i}{n_i}\right)/\hat{V}\left(\frac{h_i}{n_i}\right)\right] - \frac{k}{2},
\end{equation}
where $h_i'$s are the observed test statistics, $i = 1,2,\ldots,S$, \[\hat{E}\left(\frac{h_i}{n_i}\right) = \frac{1}{S} \sum_{i=1}^S \left(\frac{h_i}{n_i}\right)\] and \[\hat{V}\left(\frac{h_i}{n_i}\right) = \frac{1}{S}\sum_{i=1}^S \left(\frac{h_i}{n_i} - \frac{1}{S} \sum_{i=1}^S \left(\frac{h_i}{n_i}\right)\right)^2.\]

\emph{Choice of $r$ for an F-test.}
Under $H_1$, $F_i \mid \lambda_i \sim F_{k, m_i}(\lambda_i)$ and $\lambda_i \sim G\left(\frac{k}{2}+r, \frac{1}{2\tau_i^2}\right)$, $i=1,2,\ldots, S$.

\begin{equation*}
    E(F_i) = E[E(F_i \mid \lambda_i)] = \frac{m_i}{k(m_i-2)}\left(k+\left(\frac{k}{2}+r\right)2\tau_i^2\right).
\end{equation*}
This implies,
\begin{equation}
    E\left(\frac{F_i}{n_i}\right) = \frac{m_i}{k(m_i-2)}\left(\frac{k}{n_i}+\left(\frac{k}{2}+r\right)\frac{\Tilde{\omega}^2}{r}\right).
\end{equation}
Now,
\begin{eqnarray}
Var(F_i) & = & Var(E(F_i \mid \lambda_i)) + E(Var(F_i \mid \lambda_i)) \nonumber \\
& = & \left(\frac{m_i}{k(m_i-2)}\right)^2 Var(\lambda_i) + \nonumber \\
&& \left(\frac{m_i}{k(m_i-2)}\right)^2 E\left[2 \frac{(k+\lambda_i)^2+(k+2\lambda_i)(m_i-2)}{m_i-4}\right] \nonumber \\
& = & \left(\frac{m_i}{k(m_i-2)}\right)^2 \frac{1}{m_i-4} \left[2k^2 + 4k\left(\frac{k}{2}+r\right)2\tau_i^2 + \right. \nonumber \\ 
&& \left. \left(\frac{k}{2}+r\right)(2\tau_i^2)^2\left(\frac{k}{2}+r+1\right) + 2(m_i-2)k \right. \nonumber \\ 
&& \left. + 4(m_i-2)\left(\frac{k}{2}+r\right)2\tau_i^2 \right] + \left(\frac{m_i}{k(m_i-2)}\right)^2 Var(\lambda_i).\nonumber
\end{eqnarray}

Hence, assuming $k = O(1), m_i = O(n_i)$,
\begin{equation}
    Var\left(\frac{F_i}{n_i}\right) =  \left(\frac{m_i}{k(m_i-2)}\right)^2 \frac{\left(\frac{k}{2}+r\right)k^2 \Tilde{\omega}^4}{4\left(\frac{k}{2}+r-1\right)^2}+O\left(\frac{1}{n_i}\right)
\end{equation}

Ignoring the $O\left(\frac{1}{n_i}\right)$ terms, the MOM estimator of $r$ is,
\begin{equation}
    \hat{r}_{MOM} =\left[ \hat{E}^2\left(\frac{f_i}{n_i}\right)/\hat{V}\left(\frac{f_i}{n_i}\right)\right] - \frac{k}{2},
\end{equation}
where $f_i'$s are the observed test statistics, $i = 1,2,\ldots,S$, \[\hat{E}\left(\frac{f_i}{n_i}\right) = \frac{1}{S} \sum_{i=1}^S \left(\frac{f_i}{n_i}\right)\] and \[\hat{V}\left(\frac{f_i}{n_i}\right) = \frac{1}{S}\sum_{i=1}^S \left(\frac{f_i}{n_i} - \frac{1}{S} \sum_{i=1}^S \left(\frac{f_i}{n_i}\right)\right)^2.\]
 \appendix

\bibliographystyle{plainnat}
\bibliography{Statistics_And_probability_letters/els-cas-templates/cas-refs}